\documentclass[preprintnumbers]{revtex4}

\newif\ifpdf
\ifx\pdfoutput\undefined
\pdffalse % we are not running PDFLaTeX
\else
\pdfoutput=1 % we are running PDFLaTeX
\pdftrue
\fi
 
\ifpdf
\usepackage[pdftex]{graphicx}
\else
\usepackage{graphicx}
\fi
\def\bit{\begin{itemize}}                                                      
\def\eit{\end{itemize}}

 \newcommand{\note}[1]{}
\newcommand{\delhad}{\mbox{$\Delta \alpha_{\rm had}^{(5)}(M_Z)$}} 
\newcommand{\msb}{\mbox{$\overline{\rm{MS}}\ $}}                                
\newcommand{\mt}{\mbox{$m_t$}}                                                  
\newcommand{\mh}{\mbox{$M_H$}}                                                  
\newcommand{\mz}{\mbox{$M_Z$}}                                                  
\newcommand{\mw}{\mbox{$M_W$}}                                                  
                                      
\newcommand{\als}{\mbox{$\alpha_s$}}                                            
\newcommand{\suf}{\mbox{$SU(5)\ $}}

\newcommand{\skipblk}[1]{}                                                      
\def\bqa{\begin{eqnarray}}                                                      
\def\eqa{\end{eqnarray}}                                                        
\newcommand{\ee}{\mbox{$e^+ e^-$}}

\newcommand{\etal}{{\em et al., }}

\newcommand{\sto}{\mbox{$SU(2) \x U(1)\ $}}                                       
                                               
\newcommand{\x}{\mbox{$\times$}}

\newcommand{\sthto}{\mbox{$SU(3) \x SU(2) \x U(1)\ $}}                             
\newcommand{\sinn}{\mbox{$\sin^2\theta_W\,$}}                                   
                                            
\newcommand{\snu}{\mbox{$\stackrel{(-)}{\nu}$}}                                 
\newcommand{\beq}{\begin{equation}}                                             
\newcommand{\eeq}{\end{equation}}

\newcommand{\RA}{\mbox{$\rightarrow$}}

\def\mxth{\mathsurround=0pt }
\def\xversim#1#2{\lower2.pt\vbox{\baselineskip0pt \lineskip-.5pt
  \ialign{$\mxth#1\hfil##\hfil$\crcr#2\crcr\sim\crcr}}}             
\def\simgr{\mathrel{\mathpalette\xversim >}}                                    
\def\simle{\mathrel{\mathpalette\xversim <}}                                    
                                       
\begin{document}

 \ifpdf
\DeclareGraphicsExtensions{.jpg,.pdf,.mps,.png}
 \else
\DeclareGraphicsExtensions{.eps,.ps}
 \fi

% You should use BibTeX and revtex.bst for references
\bibliographystyle{revtex}

% Use the \preprint command to place your local institutional report
% number  and your conference paper identification number on the
% title page in preprint mode. Multiple \preprint commands are allowed.
\preprint{UPR 0959T}

%Title of paper
\title{Precision Electroweak Data: Phenomenological Analysis}
% Optional argument for running titles on pages
%\title[]{}

% repeat the \author .. \affiliation  etc. as needed
% \email, \thanks, \homepage, \altaffiliation all apply to the current
% author. Explanatory text should go in the []'s, actual e-mail
% address or url should go in the {}'s for \email and \homepage.
% Please use the appropriate macro for the type of information

% \affiliation command applies to all authors since the last
% \affiliation command. The \affiliation command should follow the
% other information

\author{Paul Langacker}
\email[]{pgl@electroweak.hep.upenn.edu}
%\homepage[]{Your web page}
%\thanks{}
%\altaffiliation{}
\affiliation{ School of Natural Sciences,
Institute for Advanced Study,
Princeton, NJ 08540}

\altaffiliation{Permanent address: 
Department of Physics and Astronomy,
University of Pennsylvania,
Philadelphia, PA 19104}

\date{\today}

\begin{abstract}
The precision electroweak program, including weak neutral current
(WNC), $Z$-pole, and high energy collider experiments, has been the primary
prediction and test of electroweak unification. It has established
that the standard model (SM) is correct and unique to first approximation,
establishing the gauge
principle as well as the SM gauge group and representations;
shown that the SM is correct at loop level, confirming the basic principles of
renormalizable gauge theory and allowing the successful prediction 
or constraint on $m_t$, $\alpha_s$, and $M_H$;
severely constrained new  physics at the TeV scale, with the
ideas of unification strongly favored over TeV-scale  compositeness; and
yielded precise values for the  gauge couplings, consistent with 
(supersymmetric) gauge unification.
\end{abstract}

%\maketitle must follow title, authors, abstract and \pacs
\maketitle

\section{The precision program}

The weak neutral current was a critical prediction of the electroweak
standard model (SM)~\cite{sirlinfest}. Following its discovery in 1973, there
were generations of ever more precise WNC experiments, including
pure weak $\nu N$ and $\nu e$ scattering processes,
weak-electromagnetic interference processes such as polarized
$e^{\uparrow \downarrow}D$ or $\mu N$, $\ee \RA $ (hadron or charged lepton)
cross sections and asymmetries below the $Z$ pole, and parity-violating
effects in heavy atoms (APV).  There were also  early direct observations
of the $W$ and $Z$.
The program
was supported by theoretical efforts in the
calculation of QCD and electroweak
radiative corrections; the expectations for observables
in the standard model,
large classes of extensions, and alternative models; and global
analyses of the data.
Even before the beginning of the $Z$-pole experiments at LEP and SLC in 1989,
this program had established~\cite{general}-\cite{costa}:
\begin{itemize}
\item The SM is correct to first approximation.
The four-fermion operators for $\nu q$, $\nu e$,
and $eq$ were uniquely determined,
in agreement with the standard model. 
The $W$ and $Z$ masses agreed with the expectations
of the \sto gauge group and canonical Higgs mechanism, eliminating contrived
alternative models with the
same four-fermi interactions as the standard model.
      \item Electroweak radiative corrections  were necessary  for the agreement
of theory and experiment.
        \item The weak mixing angle (in the on-shell renormalization scheme) 
was determined to be \sinn = 0.229 $\pm 0.0064$; consistency of the various
observations, including radiative corrections,  required
$m_t < 200$ GeV.
\item Theoretical uncertainties, especially in the $c$ threshold 
in deep inelastic WCC scattering,
dominated.
        \item The combination of WNC and WCC data uniquely
determined the $SU(2)$ representations of all of the known fermions,
i.e., of the $\nu_e$ and $\nu_\mu$, as well as the $L$ and
$R$ components of the $e, \ \mu, \ \tau, \ d, \, s, \, b, \ u,$ and $c$~\cite{unique}.
In particular,  the left-handed $b$ and $\tau$ were  the
lower components of $SU(2)$ doublets, implying unambiguously that the $t$ quark
and $\nu_\tau$ had to exist.
%This was independent of theoretical arguments based on
%anomaly cancellation (which could have been evaded in alternative models
%involving a vector-like third family), and of
%constraints on the $t$ mass from electroweak loops.
        \item The electroweak gauge couplings were
well-determined, allowing a detailed comparison with the gauge
unification predictions of the simplest grand unified theories (GUT).
Ordinary
    \suf was excluded (consistent with the non-observation of proton decay),
but the supersymmetric extension was allowed.
%i.e., that the data
%was ``consistent with SUSY
%GUTS and perhaps even the first harbinger of supersymmetry''~\cite{amaldi}.
        \item There were stringent limits on new physics at the TeV scale, including
additional $Z'$ bosons, exotic fermions (for which both WNC and WCC 
constraints were crucial), exotic Higgs representations,
 leptoquarks, and new four-fermion operators.
\end{itemize}

The LEP/SLC era greatly improved the precision of the electroweak program.
It allowed the differentiation between non-decoupling extensions to the
SM (such as most forms of dynamical symmetry breaking and other types
of TeV-scale compositeness), which typically predicted several
\% deviations, and decoupling extensions (such as most of the 
parameter space for supersymmetry), for which the deviations are
typically 0.1\%.

The first phase of the LEP/SLC program involved running at the $Z$
pole, $e^+ e^- \rightarrow Z \rightarrow \ell^+ \ell^-, \ \
  q \bar{q},$ and $\nu \bar{\nu}$. During the period 1989-1995 the
four LEP experiments ALEPH, DELPHI, L3, and OPAL at CERN
observed $\sim  2 \times 10^{7} Z's$. The SLD experiment at the SLC at
SLAC observed some $5 \times 10^5$ events. Despite the much lower statistics,
the SLC had the considerable advantage of a highly polarized $e^-$ beam,
with $P_{e^-} \sim$ 75\%. There were quite a few $Z$ pole observables,
including:
 \begin{itemize}
   \item The lineshape: $M_Z, \Gamma_Z,$ and the peak cross section $ \sigma$.
   \item The branching ratios for $e^+e^-,\ \mu^+ \mu^-,\ \tau^+ \tau^-, 
\ q \bar{q},\ c \bar{c},\ b \bar{b},$ and $ s \bar{s}$. One could also determine
the invisible width, $\Gamma({\rm inv})$, from which
one can derive the number 
%$\nu \bar{\nu} \Rightarrow 
$N_\nu = 2.985
\pm 0.008$ of active (weak doublet) neutrinos with 
       $m_\nu < M_Z/2$, i.e., there are only 3 conventional families with 
light neutrinos. $\Gamma({\rm inv})$ also constrains other invisible
particles, such as light sneutrinos and the light majorons associated with some 
models of neutrino mass.
   \item A number of asymmetries, including forward-backward (FB) asymmetries; 
the $\tau$ polarization, $P_\tau$;  the polarization asymmetry $A_{LR}$ associated
with $P_{e^-}$; and
mixed polarization-FB asymmetries.
    \end{itemize}
The expressions for the observables are summarized in
Appendix~\ref{P1_langacker_0702_lineshape},
and the experimental values and SM predictions in
Table~\ref{P1_langacker_0702_zpole}.
These combinations of observables could be used to isolate many
$Z$-fermion couplings, verify lepton family universality,
determine \sinn in numerous ways, and determine or constrain \mt, \als, and \mh.
 LEP and SLC simultaneously carried out other  programs,
most notably studies and tests of QCD, and heavy quark physics.

LEP~2 ran from 1995-2000, with energies gradually increasing from $\sim 140$ to $\sim 208$ GeV.
The principal electroweak results were precise measurements of the $W$ mass, as well
as its width and branching ratios (these were measured independently at the Tevatron);
a measurement of  $e^+ e^- \RA W^+ W^-$, $ZZ$, and single $W$,
as a function of center of mass (CM)
energy, which tests the cancellations between diagrams that is characteristic
of a renormalizable gauge field theory, or, equivalently, probes the triple
gauge vertices;
limits on anomalous quartic gauge vertices;
measurements of various cross sections and asymmetries for
$e^+ e^- \RA f \bar{f}$ for $f=\mu^-,\tau^-,q,b$ and $c$, in reasonable
agreement with SM predictions;
a stringent lower limit of 113.5 GeV on the Higgs mass, and even hints
of an observation at $\sim$ 115 GeV;
and searches for supersymmetric or other exotic particles.

In parallel with the LEP/SLC program, there were much more
precise ($< $ 1\%) measurements of atomic parity violation (APV) in cesium at Boulder,
along with the atomic calculations and related measurements needed for the
interpretation; precise new measurements of deep inelastic
scattering by the NuTeV collaboration at Fermilab, with
a sign-selected beam which allowed them to minimize the effects of the $c$ threshold
and reduce uncertainties to around 1\%; and few \% measurements of $\snu_\mu e$ by CHARM II
at CERN. Although the precision of these WNC processes was  lower
than the $Z$ pole measurements, they are still of considerable importance:
the $Z$ pole  experiments are blind to  types of new physics
that do not directly affect the $Z$,
such as a heavy $Z'$ if there is no $Z-Z'$ mixing,  while the WNC experiments are often very
sensitive. During the same period there were important electroweak results 
from CDF and D$\not{\! 0}$ at the Tevatron, most notably a precise value for $M_W$,
competitive with and complementary to the LEP~2 value; a direct measure of \mt,
and direct searches for  $Z'$, $W'$, exotic fermions, and supersymmetric particles.
Many of these non-$Z$ pole results are summarized in
Table~\ref{P1_langacker_0702_nonzpole}.

\begin{table} \centering
\begin{tabular}{|l|c|c|c|r|}
\hline Quantity & Group(s) & Value & Standard Model & pull \\ 
\hline
$M_Z$ \hspace{14pt}      [GeV]&     LEP     &$ 91.1876 \pm 0.0021 $&$ 91.1874 \pm 0.0021 $&$ 0.1$ \\
$\Gamma_Z$ \hspace{17pt} [GeV]&     LEP     &$  2.4952 \pm 0.0023 $&$  2.4966 \pm 0.0016 $&$-0.6$ \\
{ $\Gamma({\rm had})$\hspace{8pt}[GeV]}&
 LEP  &$  1.7444 \pm 0.0020 $&$  1.7429 \pm 0.0015 $&  ---  \\
{ $\Gamma({\rm inv})$\hspace{11pt}[MeV]}& LEP  &
$499.0    \pm 1.5    $&$501.76   \pm 0.14   $&  ---  \\
{ $\Gamma({\ell^+\ell^-})$ [MeV]}& 
   LEP     &$ 83.984  \pm 0.086  $&$ 84.019  \pm 0.027  $&  ---  \\
$\sigma_{\rm had}$ \hspace{12pt}[nb]&LEP    &$ 41.541  \pm 0.037  $&$ 41.477  \pm 0.014 
$&{ $1.7$} \\
$R_e$                         &     LEP     &$ 20.804  \pm 0.050  $&$ 20.744  \pm 0.018  $&$ 1.2$ \\
$R_\mu$                       &     LEP     &$ 20.785  \pm 0.033  $&$ 20.744  \pm 0.018  $&$ 1.2$ \\
$R_\tau$                      &     LEP     &$ 20.764  \pm 0.045  $&$ 20.790  \pm 0.018  $&$-0.6$ \\
$A_{FB} (e)$                  &     LEP     &$  0.0145 \pm 0.0025 $&$  0.01637 \pm 0.00026 $&$-0.8$ \\
$A_{FB} (\mu)$                &     LEP     &$  0.0169 \pm 0.0013 $&$                    $&$ 0.4$ \\
$A_{FB} (\tau)$               &     LEP     &$  0.0188 \pm 0.0017 $&$                    $&$ 1.4$ \\
\hline
$R_b$                         &  LEP + SLD  &$  0.21664\pm 0.00068$&$  0.21569\pm 0.00016$&$ 1.4$ \\
$R_c$                         &  LEP + SLD  &$  0.1729 \pm 0.0032 $&$  0.17230 \pm 0.00007 $&$ 0.2$ \\
%$R_{s,d}/R_{(d+u+s)}$         &     OPAL 
%&$  0.371  \pm 0.023  $&$  0.3592 \pm 0.0001 $&$ 0.5$ \\
$A_{FB} (b)$                  &     LEP     &$  0.0982 \pm 0.0017 $&$  0.1036 \pm 0.0008 $&{ $-3.2$} \\
$A_{FB} (c)$                  &     LEP     &$  0.0689 \pm 0.0035 $&$  0.0740 \pm 0.0006 $&$-1.5$ \\
$A_{FB} (s)$                  &DELPHI,OPAL&$  0.0976 \pm 0.0114 $&$  0.1037 \pm 0.0008 $&$-0.5$ \\
$A_b$                         &     SLD     &$  0.921  \pm 0.020  $&$  0.9347 \pm 0.0001 $&$-0.7$ \\
$A_c$                         &     SLD     &$  0.667  \pm 0.026  $&$  0.6681 \pm 0.0005 $&$ 0.0$ \\
$A_s$                         &     SLD     &$  0.895   \pm 0.091   $&$  0.9357 \pm 0.0001 $&$-0.4$ \\
\hline
$A_{LR}$ (hadrons)            &     SLD     &$  0.15138\pm 0.00216$&$  0.1478 \pm 0.0012 $&$ 1.7$ \\
$A_{LR}$ (leptons)            &     SLD     &$  0.1544 \pm 0.0060 $&$                    $&$ 1.1$ \\
$A_\mu$                       &     SLD     &$  0.142  \pm 0.015  $&$                    $&$-0.4$ \\
$A_\tau$                      &     SLD     &$  0.136  \pm 0.015  $&$                    $&$-0.8$ \\
%A_e (Q_{LR})$                & 
%SLD     &$  0.162  \pm 0.043  $&$                    $&$ 0.3$ \\
$A_\tau ({\cal P}_\tau)$      &     LEP     &$  0.1439 \pm 0.0041 $&$                    $&$-0.9$ \\
$A_e ({\cal P}_\tau)$         &     LEP     &$  0.1498 \pm 0.0048 $&$                    $&$ 0.4$ \\
$\bar{s}_\ell^2 (Q_{FB})$     &     LEP     &$  0.2322 \pm 0.0010 $&$  0.23143\pm 0.00015$&$ 0.8$ \\
\hline
\end{tabular}
\caption{Principal $Z$-pole observables, their experimental values, 
theoretical predictions using the SM parameters from the global best
fit~\cite{pdg01}, and pull
(difference from the prediction divided by the uncertainty).
$\Gamma({\rm had})$, $\Gamma({\rm inv})$, and $\Gamma({\ell^+\ell^-})$ are not independent,
but are included for completeness.}
\label{P1_langacker_0702_zpole}
\end{table}

\begin{table} \centering
\begin{tabular}{|l|c|c|c|r|}
\hline Quantity & Group(s) & Value & Standard Model & pull \\ 
\hline
$m_t$\hspace{8pt}[GeV]&Tevatron &$ 174.3    \pm 5.1               $&$ 175.3    \pm 4.4    $&$-0.2$ \\
$M_W$ [GeV]    &      LEP       &$  80.446  \pm 0.040             $&$  80.391  \pm 0.019  $&$ 1.4$ \\
$M_W$ [GeV]    & Tevatron,UA2 &$  80.451  \pm 0.061             $&$                     $&$ 1.0$ \\
\hline
$R^-$          &     NuTeV      &$   0.2277 \pm 0.0021 \pm 0.0007 $&$   0.2300 \pm 0.0002 $&$-1.1$ \\
$\kappa^\nu$        &     CCFR       &$   0.5820 \pm 0.0027 \pm 0.0031 $&$   0.5833 \pm 0.0004 $&$-0.3$ \\
$R^\nu$        &     CDHS       &$   0.3096 \pm 0.0033 \pm 0.0028 $&$   0.3093 \pm 0.0002 $&$ 0.1$ \\
$R^\nu$        &     CHARM      &$   0.3021 \pm 0.0031 \pm 0.0026 $&$                     $&{ $-1.7$}
\\
$R^{\bar\nu}$  &     CDHS       &$   0.384  \pm 0.016  \pm 0.007  $&$   0.3862 \pm 0.0002 $&$-0.1$ \\
$R^{\bar\nu}$  &     CHARM      &$   0.403  \pm 0.014  \pm 0.007  $&$                     $&$ 1.0$ \\
$R^{\bar\nu}$  &     CDHS 1979  &$   0.365  \pm 0.015  \pm 0.007  $&$   0.3817 \pm 0.0002 $&$-1.0$ \\
\hline
$g_V^{\nu e}$  &     CHARM II   &$  -0.035  \pm 0.017             $&$  -0.0398 \pm 0.0003 $&  ---  \\
$g_V^{\nu e}$  &      all       &$  -0.040  \pm 0.015             $&$                     $&$-0.1$ \\
$g_A^{\nu e}$  &     CHARM II   &$  -0.503  \pm 0.017             $&$  -0.5065 \pm 0.0001 $&  ---  \\
$g_A^{\nu e}$  &      all       &$  -0.507  \pm 0.014             $&$                     $&$ 0.0$ \\
\hline
$Q_W({\rm Cs})$&     Boulder    &$ -72.65   \pm 0.28\pm 0.34      $&$ -73.10   \pm 0.03   $&$ 1.0$ \\
$Q_W({\rm Tl})$&Oxford,Seattle&$-114.8    \pm 1.2 \pm 3.4       $&$-116.67    \pm 0.05    $&$ 0.5$ \\
\hline
${\Gamma (b\rightarrow s\gamma)\over \Gamma (b\rightarrow c e\nu)}$& CLEO 
           &$ 3.26^{+0.75}_{-0.68} \times 10^{-3} $&$ 3.14^{+0.17}_{-0.16}
           \times 10^{-3} $&$ 0.2$ \\
${1\over 2} (g_\mu - 2 - {\alpha\over \pi})$ &  E821 &
$4510.55 \pm 1.51 \pm 0.51$ & $4506.55 \pm 0.36 $ &2.5  \\
\hline
\end{tabular}
\caption{Recent non-$Z$-pole observables. From~\cite{pdg01}.}
\label{P1_langacker_0702_nonzpole}
\end{table}

%******************************************************************************
The LEP and (after initial difficulties) SLC programs were
remarkably successful, achieving  greater precision than
had been anticipated in the planning stages, e.g., due to better 
than expected measurements of the   beam energy
(using a clever resonant depolarization technique) and luminosity.

The effort required
the calculation of the needed electromagnetic,
electroweak,  QCD, and mixed radiative corrections
to the predictions of the SM. Careful consideration of
the competing definitions of the renormalized \sinn
was needed. 
The principal theoretical uncertainty is the hadronic
contribution \delhad \ to the running of 
$\alpha$ from its precisely known value at low energies
to the $Z$-pole, where it is needed to compare
the $Z$ mass with the asymmetries and other observables.
The radiative corrections, renormalization schemes, and
running of $\alpha$ are further discussed in
Appendix~\ref{P1_langacker_0702_radiativecorr}.
The LEP Electroweak Working Group (LEPEWWG)~\cite{LEPEWWG} 
combined the results of
the four LEP experiments, and also those of SLD and some WNC and Tevatron
results, taking proper account of
common systematic and theoretical uncertainties.
Much theoretical effort  also went into the development,
testing, and comparison of radiative corrections packages, and
into the study of how various classes of new
physics would modify the observables, and how they could
most efficiently be parametrized.

%******************************************************************************
%******************************************************************************

\section{Fits to the standard model}
\label{P1_langacker_0702_globalfits}
Global fits allow uniform theoretical treatment and exploit the
fact that the data collectively contain much more information than 
individual experiments. However, they require a careful
consideration of experimental and theoretical systematics and their correlations.
The results here are  from work with Jens Erler
for the 2001 update of  the electroweak review in the {\em Review of 
Particle Properties}~\cite{pdg01}. They incorporate
the full $Z$-pole, WNC (especially important for constraining some types of new physics),
and relevant hadron collider and LEP~2 results. The radiative corrections were
calculated with GAPP ({\em Global Analysis of Particle Properties})~\cite{GAPP}.
GAPP is fully \msb, which minimizes the mixed QCD-EW corrections
and their uncertainties and is a complement to ZFITTER~\cite{Bardin01}, 
which is on-shell.
We use a \delhad \  which is properly correlated with \als~\cite{alhad},
and also with the hadronic vacuum
polarization contribution to $g_\mu-2$~\cite{luo}.

The data are  for the most part
in excellent agreement with the SM predictions. The best fit
values for the SM parameters (as of 07/01) are,
\bqa
           M_H &=& 98^{+51}_{-35} \mbox{ GeV}, \nonumber \\
           m_t &=& 175.3  \pm 4.4  \mbox{ GeV}, \nonumber  \\
      \alpha_s &=& 0.1200 \pm 0.0028, \label{P1_langacker_0702_fitresults} \\
   \hat{s}^2_Z &=& 0.23113 \pm 0.00015, \nonumber \\
% \bar{s}^2_\ell &=& 0.23143 \pm 0.00015, \nonumber \\
%         s^2_W &=& 0.22278 \pm 0.00036 \nonumber \\
%         s^2_{M_Z} &=& 0.23105 \pm 0.00008 \nonumber \\
 \Delta \alpha_{\rm had}^{(5)}(M_Z) &=& 0.02778 \pm 0.00020. \nonumber
\eqa
\bit 

\item 
This fit included the direct (Tevatron) measurements of \mt \ and
the theoretical value of \delhad \ as constraints, but
did not include other determinations of \als \ or the LEP~2 direct limits
on \mh.

\item
The \msb value of \sinn  ($\hat{s}^2_Z$) can be translated
into  other definitions. The effective angle $\bar{s}^2_\ell = 0.23143 \pm 0.00015$
is closely related to $\hat{s}^2_Z$. The larger  uncertainty in the
on-shell $s^2_W = 0.22278 \pm 0.00036$ is due to its (somewhat artificial) dependence 
on \mh \ and \mt.
On the other hand,  the $Z$-mass definition $s^2_{M_Z} =  0.23105 \pm 0.00008$
has no \mh \ or \mt \ dependence, but the uncertainties reemerge
when comparing with other observables.

\item The best fit value $ \Delta \alpha_{\rm had}^{(5)}(M_Z) = 0.02778 \pm 0.00020$
is dominated by the theoretical input constraint $ \Delta \alpha_{\rm had}^{(5)}(M_Z) =
0.02779 \pm 0.00020$. However,
$ \Delta \alpha_{\rm had}^{(5)}(M_Z)$ can be determined from the
indirect data alone, i.e., from the relation of $M_Z$ and $M_W$ to the
other observables, and by its correlation with $g_\mu-2$.
The result,
$0.02866 \pm 0.00040$, is $\sim 1.9\sigma$ above  the theoretical value,
mainly because of $g_\mu-2$.

\item
Similarly, the value $m_t = 175.3  \pm 4.4$ GeV includes the direct Tevatron constraint
$\mt = 174.3 \pm 5.1$. However, one can determine 
$\mt = 178.1^{+10.4}_{-8.3}$ GeV from indirect data (loops) only, in excellent agreement.

\begin{figure}[h]
\centering
\includegraphics*[scale=0.6]{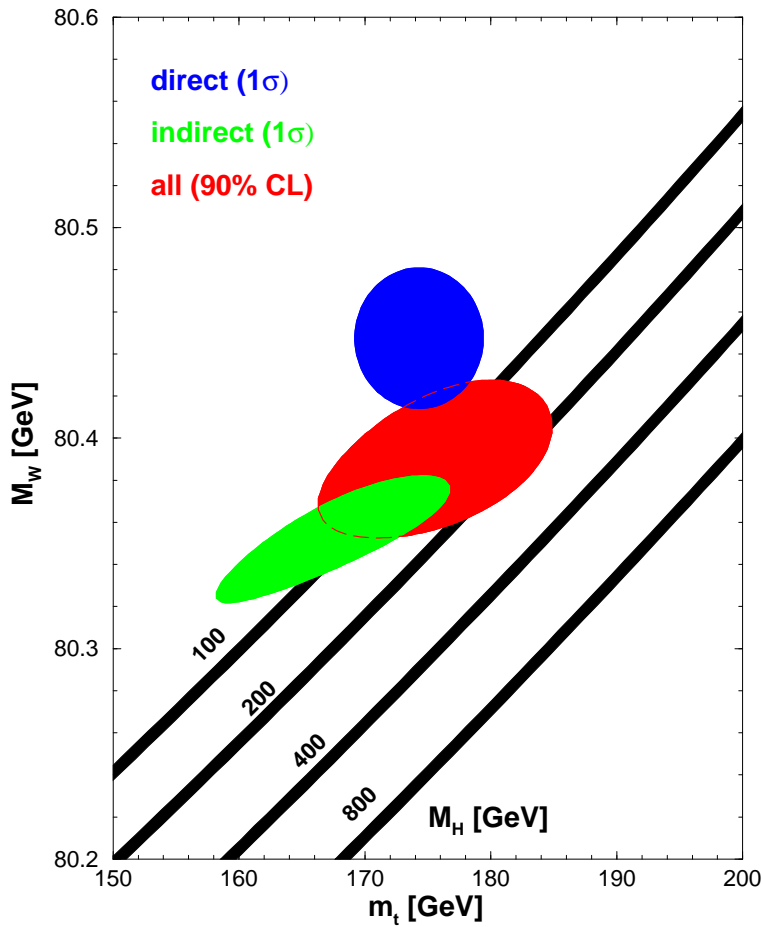}
\caption{Allowed regions in $M_W$ vs \mt \ from direct, indirect,
and combined data, compared with the standard model expectations as a function of
\mh. From~\cite{pdg01}.}
\label{P1_langacker_0702_mwmt}
\end{figure}

\begin{figure}[h]
\centering
\includegraphics*[scale=0.6]{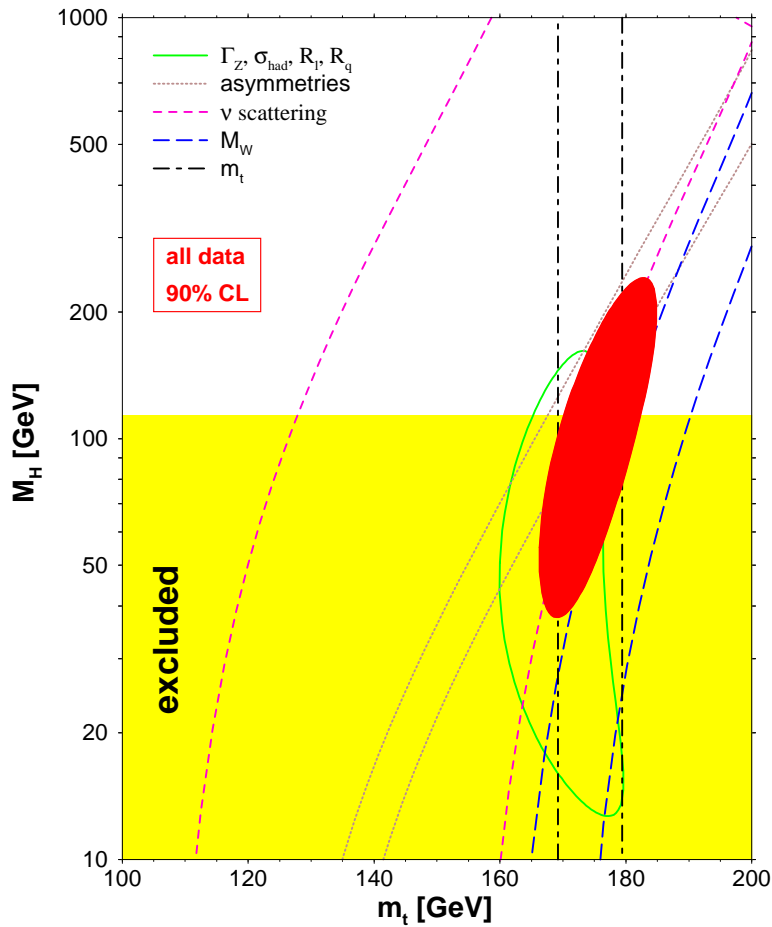}
\caption{Allowed regions in \mh \  vs \mt \ from precision data,
compared with the direct exclusion limits from LEP~2.
From~\cite{pdg01}.}
%\label{}
\end{figure}

\begin{figure}[h]
\centering
\includegraphics*[scale=0.5]{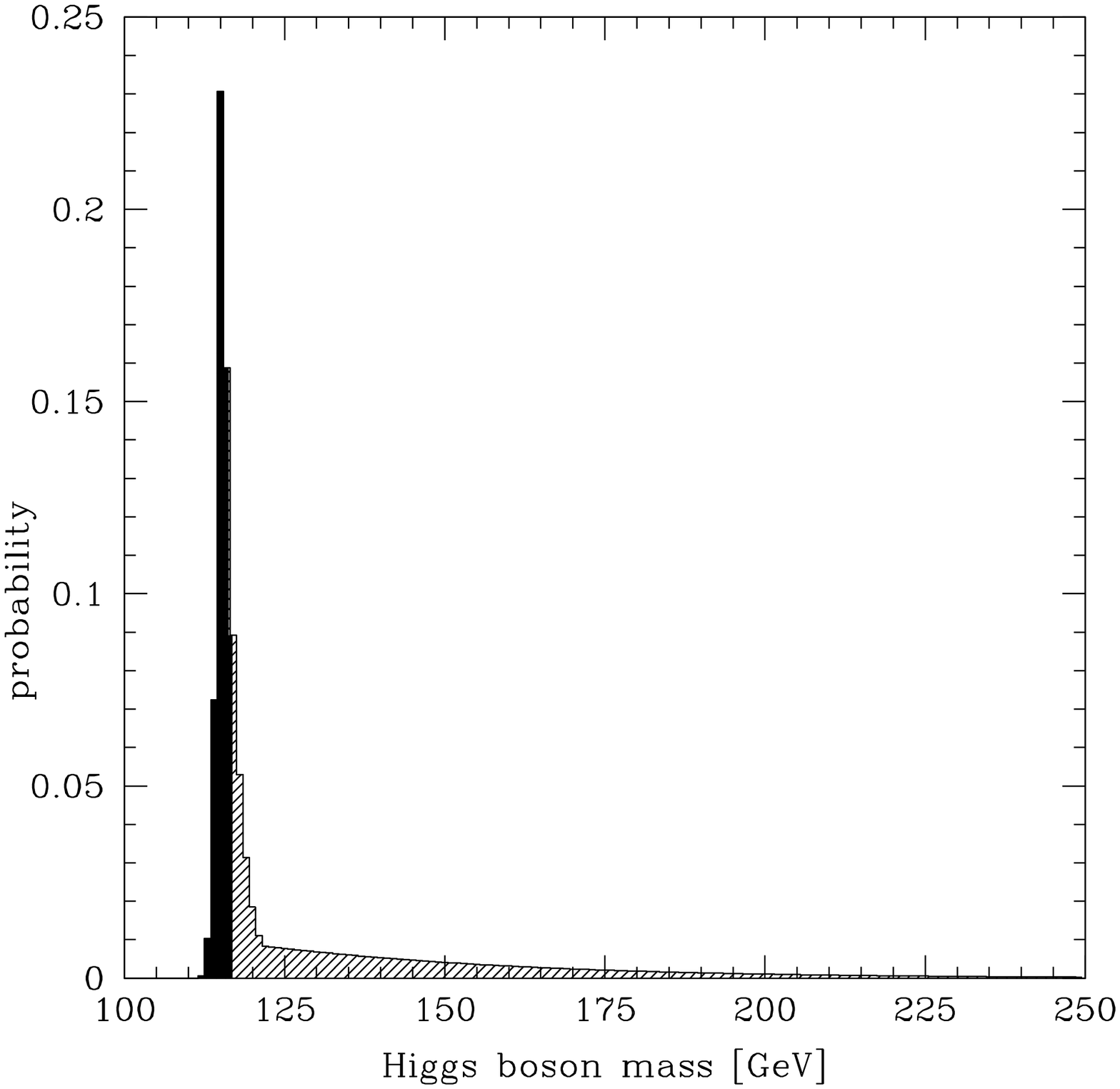}
\caption{Probability density for \mh, including direct LEP~2 data and indirect
constraints. From~\cite{erlerhiggs}.}
%\label{}
\end{figure}

\item The value \als $=0.1200 \pm 0.0028$ is consistent with  other 
determinations, e.g., from deep inelastic scattering, hadronic $\tau$
decays,  the charmonium and upsilon spectra, and
jet properties.
%The current
%PDG average (excluding the $Z$ lineshape) is 0.1182 $\pm$ 0.0013.

 \item
The central value of the 
Higgs mass prediction from the fit, \mh $= 98^{+51}_{-35}$ GeV,
is below the direct lower limit from LEP~2 of $\simgr 113.5$ GeV,
or their candidate events at 115 GeV, but consistent at the $1\sigma$ level. 
Including the direct LEP~2
likelihood function~\cite{erlerhiggs,degrassi}  along with the indirect data, one obtains
 $\mh < 199$ GeV at 95\%. Even though \mh \ only enters the precision
data logarithmically (as opposed to the quadratic \mt \ dependence),
the constraints are significant. They are also fairly
robust to most, but not all, types of new physics.
(The limit on \mh \ disappears if one allows an arbitrarily large negative
$S$ parameter and/or
a large positive $T$
(section~\ref{P1_langacker_0702_newphysics}), but most extensions of the
SM yield $S > 0$.)
One caveat is that $M_W$ and $A_{LR}$ especially favor rather low values
of \mh,  while $A_{FB}(b)$, which deviates by 3.2$\sigma$
from the SM (see below), compensates by favoring a high value~\cite{Chanowitz}.
The predicted range should be compared with the
theoretically expected range in the standard model: 
115 GeV $\simle \mh \simle$ 750 GeV, where the lower (upper) limit
is from vacuum stability (triviality). On the other hand,
the MSSM predicts $\mh \simle 130$ GeV, while the limit increases to
around 150 GeV in extensions of the MSSM.

\item 
The results in (\ref{P1_langacker_0702_fitresults})
are in excellent agreement with those
of the LEPEWWG~\cite{LEPEWWG}
up to well-understood effects~\cite{pdg01},
such as more extensive
WNC inputs and small differences in higher order terms
and \delhad,  despite the different
renormalization schemes used.
The LEPEWWG obtains:
 $\bar{s}^2_\ell = 0.23142 \pm 0.00014$; \
 $\alpha_s = 0.118 \pm 0.003$; \
$m_t = 175.7^{+4.4}_{-4.3}$ GeV; and
$M_H =98^{+58}_{-38}$ GeV.

\item 
The most significant deviation from the SM is in the
forward-backward asymmetry into $b$ quarks,
$A_{FB}(b)$, which is 3.2$\sigma$ below the prediction.
If not just a statistical fluctuation or systematic problem, this could
be a hint of new physics.  However, any such effect should not contribute
too much to  $R_b$, which is consistent with the SM.
The size of the deviation suggests
a tree level effect, such as the mixing of $b_{L,R}$ with exotic
quarks~\cite{general,choudhury}.
The most recent LEP results on $M_W$ have moved slightly above the SM
prediction (table
(\ref{P1_langacker_0702_nonzpole}) and figure
(\ref{P1_langacker_0702_mwmt})), but even when combined
with the Tevatron results (also a bit high) this is only a 1.5$\sigma$
effect. The muon magnetic moment $g_\mu -2$ result could point towards
new physics, but there are still significant hadronic uncertainties.
Within the SM fits, the only affect is the correlation of the
theoretical value with \delhad,
which lowers the \mh \ prediction by $\sim$ 5 GeV~\cite{luo}.

 \begin{figure}  \centering
\includegraphics*[scale=0.5]{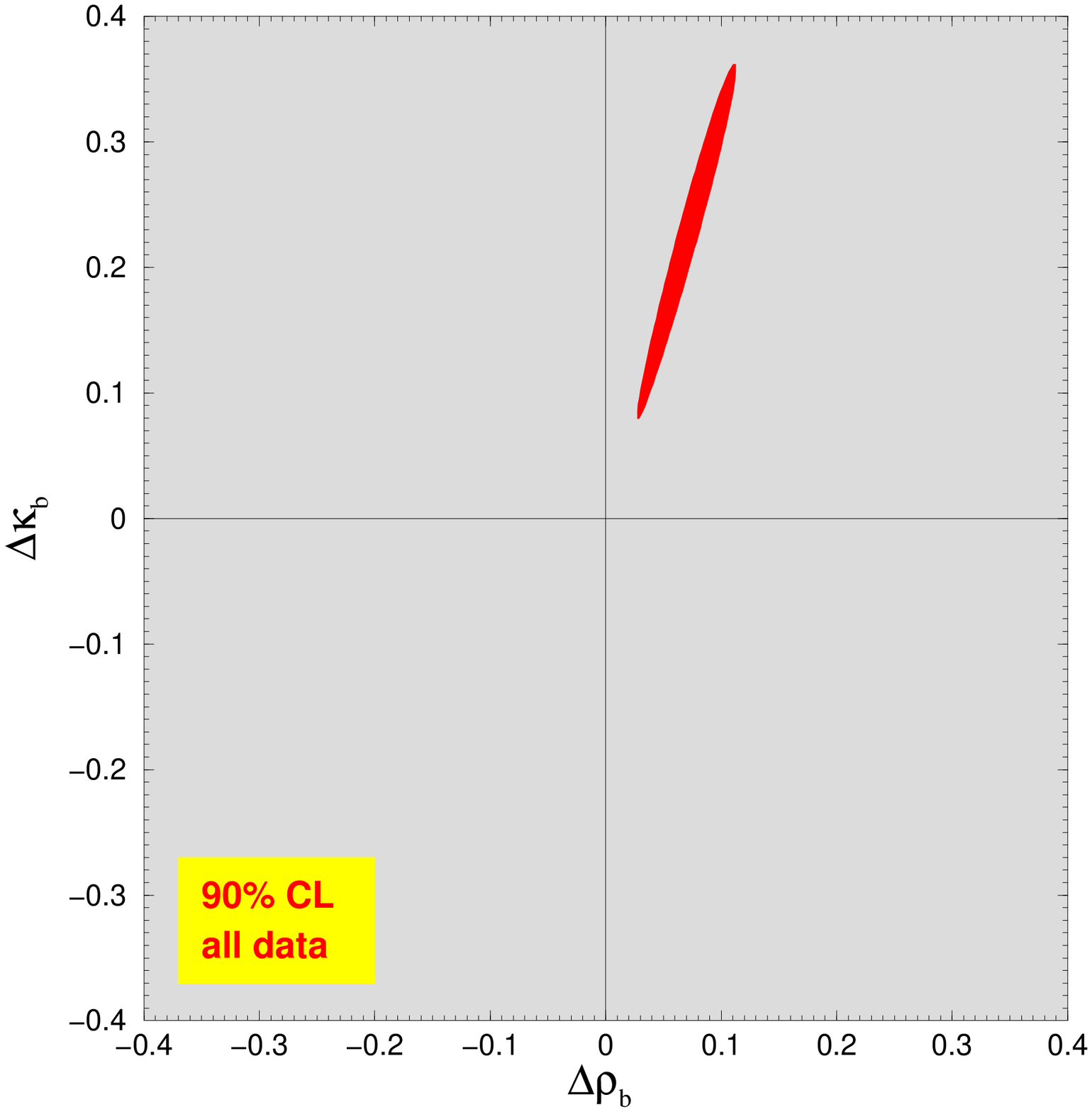}
 \caption{Regions in new physics parameters suggested by $A_{FB}(b)$
 and $R_b$. A very large ($O$(20\%)) vertex correction
 $\Delta \kappa_b$ would be required to account for the data 
 by loop effects. Courtesy of Jens Erler.}
 \label{P1_langacker_0702_zbb}
 \end{figure}

\eit

%******************************************************************************
\section{Beyond the standard model}
\label{P1_langacker_0702_newphysics}
The standard model (\sthto plus general
relativity), extended to include neutrino mass,
is the correct description of nature to first approximation down 
to $10^{-16}$ cm. However, nobody thinks that the SM is the ultimate description
of nature. It has some 28 free parameters; has a complicated gauge group and
representations; does not explain charge quantization, the fermion families,
or their masses and mixings; has several notorious fine tunings
associated with the Higgs mass, the strong CP parameter, and
the cosmological constant; and does not incorporate quantum gravity.

Many types of possible TeV scale physics are constrained by the
precision data. For example,

\bit
  \item
$S, T,$ and $ U$ parametrize  new physics sources which only affect the gauge
propagators, as well as Higgs triplets, etc.  One expects $T \ne 0$, usually positive and
often of order
unity,  from nondegenerate heavy
fermion or scalar doublets, 
while new chiral fermions (e.g., in extended technicolor (ETC)), lead to $S \ne 0$,
again usually positive and often of order unity.
The current global fit result is~\cite{pdg01}
 \bqa  S &=& -0.03 \pm 0.11 (-0.08),  \nonumber  \\
T &=& -0.02 \pm 0.13 (+0.09), \label{P1_langacker_0702_stu}  \\
U &=& 0.24 \pm 0.13 (+0.01) \nonumber 
\eqa
for $M_H = 115 \ (300)$ GeV. (We use a definition in which $S$, $T$, and $U$ 
are exactly zero in 
the SM.) The value of $S$ would be $2/3\pi$ for a heavy degenerate ordinary
or mirror family, which is therefore excluded at 99.8\%. Equivalently,
the number of families is $N_{\rm fam} = 2.97 \pm 0.30$.
This result assumes $T=U=0$, and therefore that any new families are
degenerate. This restriction can be relaxed by allowing $T \ne 0$,
yielding the somewhat weaker constraint $N_{\rm fam} = 3.27 \pm 0.45 $
for $T = 0.10 \pm 0.11$.
This is
complementary to the lineshape result $N_\nu = 2.985 \pm 0.008 $, which only applies for
 $\nu$'s
lighter than $\sim M_Z/2$. $S$ also eliminates many QCD-like ETC models.
$T$ is equivalent to the $\rho_0$ parameter~\cite{general}, which
is defined to be exactly unity in the SM. For $S = U = 0$, one obtains
 $\rho_0  \sim 1 + \alpha T = 1.0012^{+0.0023}_{-0.0014}$, with the SM
fit value for \mh \ increasing to $M_H = 211^{+814}_{-139}$ GeV.

\begin{figure}[h]
\centering
\includegraphics*[scale=0.6]{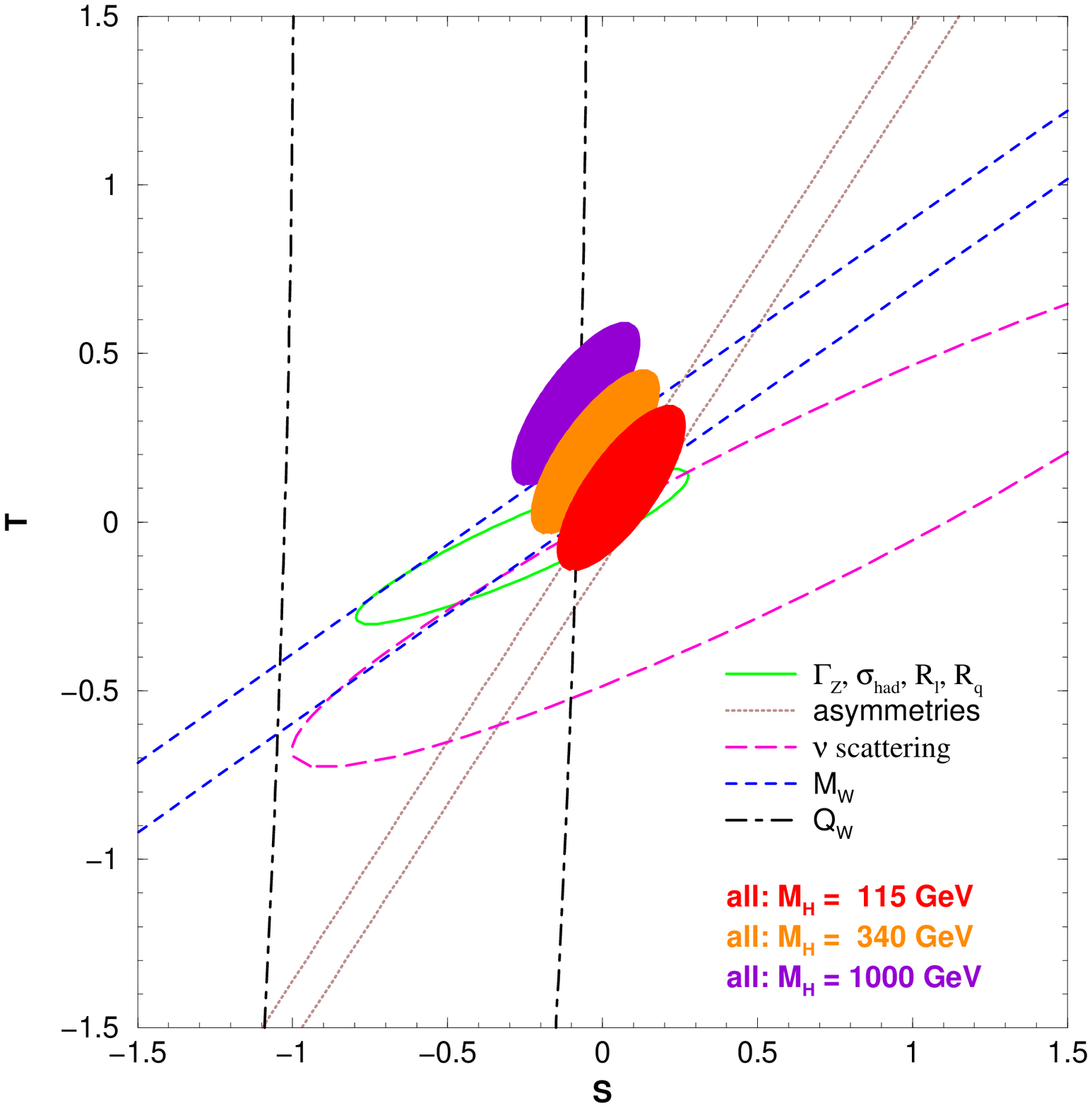}
\caption{Allowed regions in $S$ vs $T$.
From~\cite{pdg01}.}
%\label{}
\end{figure}

  \item 
Supersymmetry: in the 
decoupling limit, in which the sparticles are heavier than
$ \simgr 200-300$ GeV, there is little effect on the precision
observables, other than that there is necessarily 
  a light SM-like Higgs, consistent with the data. There is little
improvement on the SM fit, and in fact one can somewhat constrain
the supersymmetry breaking parameters~\cite{susy}.

\item Heavy $Z'$ bosons are predicted by many 
grand unified  and string theories. Limits on the $Z'$ mass
are model dependent, but are typically  around $M_{Z'} > 500-800 $ GeV 
from indirect constraints from WNC and  LEP~2 data, with comparable
limits from direct searches at the Tevatron. $Z$-pole data
severely constrains the $Z-Z'$ mixing, typically
 $|\theta_{Z-Z'}| < {\rm few} \times 10^{-3}$.

\item Gauge unification is predicted in GUTs and some string theories.
The simplest non-supersymmetric unification is excluded by
the precision data. For the MSSM, and assuming 
no new thresholds between 1 TeV and the unification scale, one
can use the precisely known $\alpha$ and $\hat{s}^2_Z$
to predict $\als = 0.130 \pm 0.010$ and a unification scale 
$M_G \sim 3 \times 10^{16}$ GeV~\cite{polonsky}. The \als \ uncertainties are 
mainly theoretical, from the TeV and GUT thresholds, etc.
\als \ is high compared to the experimental value, but barely consistent 
given the uncertainties. $M_G$ is reasonable for a GUT (and
is consistent with simple seesaw models of neutrino mass),
but is somewhat below the expectations $\sim 5 \times 10^{17}$ GeV of the simplest
perturbative heterotic string models. However, this is only a
10\% effect in the appropriate variable
$\ln M_G$. The new exotic particles often present in such models
(or higher Ka\v c-Moody levels) can easily shift the $\ln M_G$
and \als \ predictions significantly, so the problem is really
why the gauge unification works so well.
It is always possible that the apparent success is accidental
(cf., the discovery of Pluto).
\eit

\section{Conclusions}

The precision $Z$-pole, LEP~2, WNC, and Tevatron experiments have
successfully tested the SM at the 0.1\% level, including electroweak loops, thus
confirming the gauge principle,
SM   group, representations, and the
basic structure of renormalizable field theory.
The standard model parameters $\sin^2 \theta_W$, $m_t$, and $\alpha_s$
were precisely determined.
In fact, \mt \ was successfully predicted from its indirect loop effects prior
to the direct discovery at the Tevatron, while the indirect value of \als,
mainly from the $Z$-lineshape, agreed with more direct QCD determinations.
Similarly, \delhad \ and $ M_H$ were constrained.
The indirect (loop) effects implied $M_H \simle 191$ GeV, while direct
searches at LEP~2 yielded $M_H > 113.5 $ GeV, with a hint of a signal at 115 GeV.
This range is consistent with, but does not prove, 
the expectations of the supersymmetric
extension of the SM (MSSM), which predicts a light SM-like Higgs for much of
its parameter space. The agreement of the data with the SM imposes
a severe constraint on possible new physics at the TeV scale,
and points  towards decoupling theories (such as most versions of
supersymmetry and unification), which typically lead to 0.1\% effects,
rather than TeV-scale compositeness (e.g., dynamical symmetry breaking
or composite fermions), which usually imply  deviations of several \% (and often
large flavor changing neutral currents). 
Finally, the precisely measured gauge couplings were consistent with the
simplest form of grand unification if the SM is extended to the MSSM.

Although the $Z$-pole program has ended for the time being, there are
prospects for future programs using the Giga-$Z$ option at TESLA or possible
other linear colliders, which
might yield a factor $10^2$ more events. This would 
enormously improve the sensitivity~\cite{giga}, but would also require a large
theoretical effort to improve the radiative correction 
calculations.

\appendix

%%******************************************************************************

\section{The $Z$ Lineshape and Asymmetries}
\label{P1_langacker_0702_lineshape}

The $Z$ lineshape measurements determine the cross section 
$e^+ e^- \RA f \bar{f}$ for $f = e, \mu, \tau, s, b, c,$ or hadrons
as a function of $s = E_{CM}^2$. To lowest order,
\beq
\sigma_f(s) \sim \sigma_f \frac{s \Gamma^2_Z}{
  \left(s - M_Z^2\right)^2 + \frac{s^2 \Gamma_Z^2}{M_Z^2}},
\label{P1_langacker_0702_sigma}
\eeq
where
significant initial state   radiative corrections are not displayed.

The peak cross section $\sigma_f$ is related to the $Z$ mass and
partial widths by
\beq \sigma_f = \frac{12 \pi}{M_Z^2}  \ \ 
    \frac{\Gamma(e^+ e^-) \Gamma(f \bar{f})}{\Gamma_Z^2}. 
\eeq
The widths are expressed in terms of the effective $Z f \bar{f}$ vector
and axial couplings $\bar{g}_{V,Af} $ by
\beq
\Gamma(f\bar{f}) \sim \frac{C_f G_F M_Z^3}{6 \sqrt{2} \pi}
  \left[ |\bar{g}_{Vf}|^2 + |\bar{g}_{Af}|^2 \right],
\label{P1_langacker_0702_width}
\eeq
where $C_\ell = 1$ and $ \ C_q = 3$.
Electroweak radiative corrections are absorbed into the $\bar{g}_{V,Af}$.
There are fermion mass, QED, and QCD corrections to
(\ref{P1_langacker_0702_width}).

The effective couplings in
(\ref{P1_langacker_0702_width}) are defined in the SM by
\beq \bar{g}_{Af} = \sqrt{\rho_f} t_{3f}, \ \ \ \
\bar{g}_{Vf} = \sqrt{\rho_f} \left[ t_{3f} - 2 \bar{s}^2_f q_f \right],
\eeq
%\begin{eqnarray} \bar{g}_{Af} &=& \sqrt{\rho_f} t_{3f} \nonumber \\
%\bar{g}_{Vf} &=& \sqrt{\rho_f} \left[ t_{3f} - 2 \bar{s}^2_f q_f \right]
%\nonumber
%\end{eqnarray}
where $q_f$ is the electric charge and $t_{3f}$ is the weak isospin
of fermion $f$, and $\bar{s}^2_f$ is the effective weak
angle. It is related by ($f$-dependent)
vertex corrections to the on-shell or \msb \ definitions of
\sinn by
\beq \bar{s}^2_f = \kappa_f s^2_W \ \ ({\rm on-shell}) 
=
\hat{\kappa}_f \hat{s}^2_Z  \ \ (\msb). \eeq
$\rho_f-1, \  \kappa_f-1 ,$ and $\hat{\kappa}_f -1 $
are electroweak corrections. For $f=e$ and the known ranges for \mt \ and \mh,
$\bar{s}^2_e \sim \hat{s}^2_Z + 0.00029$.

It is convenient to define the ratios
\beq
R_{q_i}  \equiv \frac{\Gamma(q_i \bar{q}_i)}{\Gamma({\rm had})},\ \  \
R_{\ell_i}  \equiv \frac{\Gamma({\rm had})}{\Gamma(\ell_i \bar{\ell}_i)},
\label{P1_langacker_0702_ratios}
\eeq
which isolate the weak vertices (including the effects
of \als \ for $R_{\ell_i}$).
In (\ref{P1_langacker_0702_ratios})
$q_i = b, c, s$; $\ell_i = e, \mu, \tau$; and
$\Gamma({\rm had})$ is the width into hadrons. The data are
consistent with lepton universality, i.e., with
$R_e = R_\mu = R_\tau  \equiv R_\ell$.
The partial width into neutrinos or other invisible states
is defined by $
 \Gamma({\rm inv}) = \Gamma_Z - \Gamma({\rm had}) - \sum_i
\Gamma(\ell_i \bar{\ell}_i ),$ where $\Gamma_Z$ is obtained from the
width of the cross section and the others from the peak heights.
This allows the determination of the number of  neutrinos 
%by $\Gamma({\rm inv})/\Gamma(\ell \bar{\ell}) 
%\equiv N_\nu \Gamma(\nu \bar{\nu})/\Gamma(\ell \bar{\ell})$, 
by $\Gamma({\rm inv})
\equiv N_\nu \Gamma(\nu \bar{\nu})$, 
where $\Gamma(\nu \bar{\nu})$ is the partial
width into a single neutrino flavor.
It has become conventional to work with the parameters
$M_Z, \Gamma_Z, \sigma_{\rm had}, R_\ell, R_b, R_c$, for which the correlations
are relatively small (but still must be included).

The experimenters have generally presented the Born asymmetries, $A^0$,
for which the off-pole, $\gamma$ exchange, $P_{e^-}$, and (small)
box effects have been removed from the data.
Important asymmetries include:
\bqa
{\rm forward-backward:} & & \ \ A^{0f}_{FB} \simeq \frac{3}{4} A_e A_f, \nonumber \\
\tau \ \ {\rm polarization:} & & \ \ 
P_\tau^0 = - \frac{A_\tau + A_e \frac{2 z}{1+z^2}}{1
  + A_\tau A_e \frac{2 z}{1+z^2}}, \label{P1_langacker_0702_asymmetries}
 \\ 
e^- {\rm polarization (SLD):} & & \ \ A^0_{LR} = A_e,  \nonumber \\
 {\rm mixed \ \ (SLD):}  & & \ \ 
A^{0FB}_{LR} = \frac{3}{4} A_f. \nonumber
\eqa
The LEP experiments also measure a hadronic forward-backward charge asymmetry $Q_{FB}$.
In (\ref{P1_langacker_0702_asymmetries}), $A_f$ is defined as the ratio
\beq
 A_f \equiv  \frac{2 \bar{g}_{Vf}  \bar{g}_{Af}}{
\bar{g}_{Vf}^2 + \bar{g}_{Af}^2}
\eeq
for fermion $f$.
The forward-backward asymmetries into leptons allow
another (successful) test of lepton family universality, by
$A^{0e}_{FB} = A^{0\mu}_{FB} = A^{0\tau}_{FB} 
\equiv A^{0\ell}_{FB}$. In the $\tau$ polarization, 
$z = \cos \theta$, where $\theta $ is the  scattering angle.
The SLD polarization asymmetry $A^0_{LR}$ for hadrons (or leptons)
projects out the initial electron couplings. It is especially
sensitive to \sinn because it is linear in the small $\bar{g}_{Ve}$,
while the leptonic $A^{0\ell}_{FB}$ are quadratic. The mixed polarization-FB
asymmetry $A^{0FB}_{LR}$ projects out the final fermion coupling.

\section{Radiative Corrections}
\label{P1_langacker_0702_radiativecorr}
The data are sufficiently precise that one must include
high-order radiative corrections, including 
the dominant two-loop electroweak ($\alpha^2 m_t^4, \ \alpha^2 m_t^2$),
dominant 3 loop QCD (and 4 loop estimate),
dominant 3 loop mixed QCD-EW, and 2 loop $\alpha \alpha_s$ vertex corrections.

In including EW corrections, one must choose a definition
of the renormalized \sinn. There are several popular choices,
which are equivalent at tree-level, but differ by finite
(\mt \ and \mh \ dependent) terms at higher order. These include
  \begin{itemize}
   \item  On shell: $s^2_W \equiv 1 - \frac{M_W^2}{M_Z^2},$
   \item  $Z$ mass: $s^2_{M_Z} \left(1 - s_{M_Z}^2 \right) \equiv
\frac{\pi \alpha (M_Z)}{\sqrt{2} G_F M_Z^2},$
   \item  \msb: $\hat{s}^2_Z \equiv \frac{\hat{g}'^2 (M_Z)}{\hat{g}'^2 (M_Z) +
\hat{g}^2 (M_Z)},$
   \item   Effective ($Z$-pole):    $\bar{s}^2_f \equiv \frac{1}{4}
  \left( 1 - \frac{\bar{g}_{Vf}}{\bar{g}_{Af}}  \right).$
  \end{itemize}
The first two are {\it defined} in terms of the $Z$ and $W$ masses;
the \msb \
from the renormalized couplings $\hat{g}$, $\hat{g}'$; and
the effective from the observed  vertices. Of course, each can be
determined experimentally from any observable, given the appropriate
SM expressions.
$s^2_W$ is especially simple conceptually, but the value extracted
from $Z$ pole observables has a large \mt \ and \mh \ dependence. It
(along with $s^2_{M_Z}$) is also awkward in the presence of any type of
new physics 
which shifts the values of the physical boson masses.
The $Z$-pole $\bar{s}^2_f$ depends on the fermion $f$ in the final state.
The \msb \ definition is especially useful for comparing with
theoretical predictions and for describing non $Z$-pole experiments.
The values of
$\hat{s}^2_Z$ and $\bar{s}^2_f$ are less sensitive to most types
of new physics than the on-shell definitions.
The advantages and drawbacks of each scheme are discussed in more detail
in~\cite{sirlinfest,general}.

The expressions for \mw \ and \mz \ in the on-shell and \msb \ schemes
are
\beq 
M_W^2 = \frac{\left( \pi \alpha/\sqrt{2} G_F \right)}{
s_W^2 (1 - \Delta r)} = 
\frac{\left( \pi \alpha/\sqrt{2} G_F \right)}{
 \hat{s}^2_Z (1 - \Delta \hat{r}_W)}
\eeq
and
\beq
M_Z^2 = \frac{M_W^2}{c^2_W} = \frac{M_W^2}{
  \hat{\rho} \hat{c}^2_Z},
\eeq
where the other renormalized parameters are the fine structure constant $\alpha$
(from QED) and the Fermi constant $G_F$, defined in terms of the $\mu$ lifetime.
$\Delta r$, $\Delta \hat{r}_W$, and $\hat{\rho} - 1$ collect
the radiative corrections involving $\mu$ decay, \mw, \mz, and the running
of $\alpha$ up to the $Z$ pole. In \msb, $\Delta \hat{r}_W $
has only weak \mt \ and \mh \ dependence, and is dominated
by the running of $\alpha$, i.e, 
$\Delta \hat{r}_W \sim \Delta \alpha + \cdots  \sim 0.066 + \cdots$.
In contrast, the on-shell $\Delta r$ has an additional
large (quadratic) \mt \ dependence, which results in
a large sensitivity of the observed value of $s_W^2$ to \mt. The 
\msb \ scheme isolates the large effects in the explicit
parameter
$\hat{\rho} \sim 1 + \frac{3 G_F \hat{m}^2_t}{8 \sqrt{2} \pi^2} + \cdots$.
The various definitions are related by (\mt \ and \mh \ dependent)
form factors $\kappa$, e.g., $\bar{s}^2_f = \kappa_f s^2_W = 
\hat{\kappa}_f \hat{s}^2_Z$.
%For $f=e$ and the experimental \mt, \mh,
%one obtains $\bar{s}^2_e \sim \hat{s}^2_Z + 0.00029$.

The \msb \ weak angle $\hat{s}^2_Z$ can be obtained cleanly from
the weak asymmetries. Comparison with \mz \ and \mw \ is important
for constraining \mh \ and new physics. The largest theory uncertainty in the
$M_Z-\hat{s}^2_Z$ relation is the hadronic
contribution to the running of $\alpha$ from
its precisely known value $\alpha^{-1} \sim 137.036 $ at low energies,
to the electroweak scale, where one expects
$\alpha^{-1}(M_Z) \sim \hat{\alpha}^{-1}(M_Z) + 0.99 \sim 129$.
($\hat{\alpha}$ refers to the \msb \ scheme.)
There is a related uncertainty in the hadronic vacuum polarization
contribution to the
anomalous magnetic moment of the muon.
More explicitly, one can define $\Delta \alpha$ by
\beq
\alpha(M_Z^2) = \frac{\alpha}{1-\Delta \alpha}.
\eeq
Then,
\beq \Delta \alpha = \Delta \alpha_\ell + \Delta \alpha_t +
\Delta \alpha^{(5)}_{\rm had}
\sim  0.031497 - 0.000070   + \Delta \alpha^{(5)}_{\rm had}. \eeq
%\bqa \Delta \alpha &=& \Delta \alpha_\ell + \Delta \alpha_t +
%\Delta \alpha^{(5)}_{\rm had}  \nonumber \\
%  &\sim & 0.031497 - 0.000070   + \Delta \alpha^{(5)}_{\rm had}. \nonumber \eqa
The  leptonic and $t$ loops are reliably calculated
in perturbation theory, but not $ \Delta \alpha^{(5)}_{\rm had}$ from the
lighter quarks. $ \Delta \alpha^{(5)}_{\rm had}$ can be expressed
by a dispersion integral involving $R_{\rm had}$ (the cross section
for $e^+ e^ - \RA$ hadrons relative to  $e^ + e^- \RA \mu^+ \mu^-$).
Until recently, most calculations  were data driven, using experimental values for
$R_{\rm had}$ up to CM energies $\sim$ 40 GeV, with perturbative
QCD (PQCD) at higher energies. However, there are  significant experimental uncertainties
(and some discrepancies) in the low energy data. A number of recent
studies have argued that one could reliably use a combination
of theoretical estimates using PQCD and such non-perturbative techniques as
sum rules and operator product expansions down to $\sim$ 1.8 GeV,
leading to lower uncertainties. The on-shell evaluations use the new resonance 
data from BES~\cite{BES} as further input.
The recent estimates, which are in very good agreement, are
summarized in~\cite{pdg01}. 
One can also determine $ \Delta \alpha^{(5)}_{\rm had}$
directly from the precision fits (Section~\ref{P1_langacker_0702_globalfits}).

\begin{acknowledgments}
This work was supported by the W. M. Keck Foundation as a 
Keck Visiting Professor at the Institute for Advanced Study, by the 
Monell Foundation, and by
the U.S. Department of Energy grant DOE-EY-76-02-3071.
It is a pleasure to thank Jens Erler for his collaboration.
\end{acknowledgments}

%******************************************************************************
\end{document}